\documentclass[a4paper]{panl}
\usepackage{cite}
\usepackage{wrapfig}
\usepackage{graphicx}
\usepackage{amssymb}
\usepackage{amsfonts}
\usepackage{amsmath}
\usepackage{longtable}
\usepackage{rotating}
\usepackage{lscape}
\usepackage{epsfig}
\usepackage{multirow}
\usepackage{float}

\makeatletter
\@addtoreset{equation}{section}
\makeatother

\originalTeX
\begin{document}

\title{History of Lattice Field Theory \\ from a Statistical Perspective}
\maketitle
\authors{W.\,Bietenholz\,\footnote{E-mail: wolbi@nucleares.unam.mx}},
\setcounter{footnote}{0}
\from{Instituto de Ciencias Nucleares\\
  Universidad Nacional Aut\'{o}noma de M\'{e}xico\\
  A.P. 70-543, C.P. 04510 Ciudad de M\'{e}xico, Mexico}
\begin{abstract}
Researchers working in lattice field theory constitute an
established community since the early 1990s, and around the same
time the online open-access e-print repository {\tt arXiv} was
created. The fact that this field has a specific {\tt arXiv} section,
{\tt hep-lat}, which is comprehensively used,
provides a unique opportunity for a statistical study of
its evolution over the last three decades.
We present data for the number of entries, $E$, published papers, $P$,
and citations, $C$, in total and separated by nations. We compare them
to six other {\tt arXiv} sections
({\tt hep-ph, hep-th, gr-qc, nucl-th, quant-ph, cond-mat})
and to two socio-economic indices of the nations involved:
the Gross Domestic Product (GDP) and the Education Index (EI).
We present rankings, which are based either on the Hirsch Index H,
or on the linear combination $\Sigma = E + P + 0.05 C$. We consider
both extensive and intensive national statistics, {\it i.e.}\
absolute and relative to the population or to the GDP.\\ \ \\ 
\vspace{0.2cm}

\end{abstract}
\vspace*{6pt}


\label{sec:intro}
\section{Lattice Field Theory}

The conceptual basis of lattice field theory was elaborated in the
1970s and 1980s. There is no doubt that Ken Wilson played a key role,
{\it e.g.}\ Ref.\ \cite{Wilson} had a major impact, but we also
mention unpublished work by Jan Smit.
Independent early activities in the Soviet Union are reviewed
by Alexander Polyakov \cite{Polyakov}; he gives particular
importance to Vadim Berezinskii's work.

Since the early 1990s, lattice physicists form an established,
intercontinental community, where the annual Lattice Conference
is standard as a point of orientation. It started as a small
workshop in 1982 \cite{Urs}, in 1992 it already involved
145 contributions, and nowadays this number is around $500$.
Unlike other conference series, all participants are
allowed to present a talk or poster. It was cancelled in 2020,
held online in 2021, and in 2023 it took for the
first time a hybrid format, which will hopefully persist.

So far, there are seven textbooks fully devoted to lattice field
theory \cite{books}.
Usually lattice simulations --- or more generally: computational
physics --- is considered as a branch of theoretical physics, but
occasionally it is also viewed as a third line of research, in
addition to (or in between) theory and experiment. For instance,
like real experiments, also the ``numerical experiments'' on the
lattice lead to non-perturbative results with statistical and
systematic errors.

The methodology refers to the functional integral formulation of
Quantum Field Theory: the lattice structure implements an UV
regularization, and the use of Euclidean space-time provides a
link to Statistical Mechanics. One then applies Monte Carlo techniques
to generate field configurations, let's call them $[\Phi]$, with
probability $p[\Phi] \propto \exp(-S_{\rm E}[\Phi])$ (where $S_{\rm E}$
is the model's Euclidean action, which we assume to be real
positive). Summing over a large set of configurations yields
non-perturbative results for expectation values of observables,
in particular $n$-point functions.

Gauge invariance holds by construction at the regularized level,
and the use of compact link variables (in the gauge group, rather
than the algebra) avoids the necessity of gauge fixing. Fermion
fields are given in terms of Grassmann variables. Usually the actions
are bilinear in $\bar \Psi$ and $\Psi$, then the Grassmann integral
leads to the fermion determinant, so the computer does not need to
deal explicitly with Grassmann fields. Still, in lattice QCD
there are typically millions of components, such that the frequent
computation of the fermion determinant (which depends to the gauge
configuration) is tedious: in the 20th century, it was often just set
to a constant
(``quenched approximation''), which introduces systematic errors
of ${\cal O}(10)\,\%$ in the hadron spectrum.

Nowadays, we take the fermion determinant into account, {\it i.e.}\
we deal with dynamical quarks, hence sea quarks are included,
and the lattice results for the light hadron spectrum agree
with experiment to percent-level. As an example, Figure \ref{hadspec}
shows results by the QCDSF-UKQCD Collaboration \cite{QCD},
with members from the Germany, Japan, Mexico, Russia and the UK.

\begin{figure}[t]
\vspace{-5mm}
\begin{center}
\includegraphics[width=83mm]{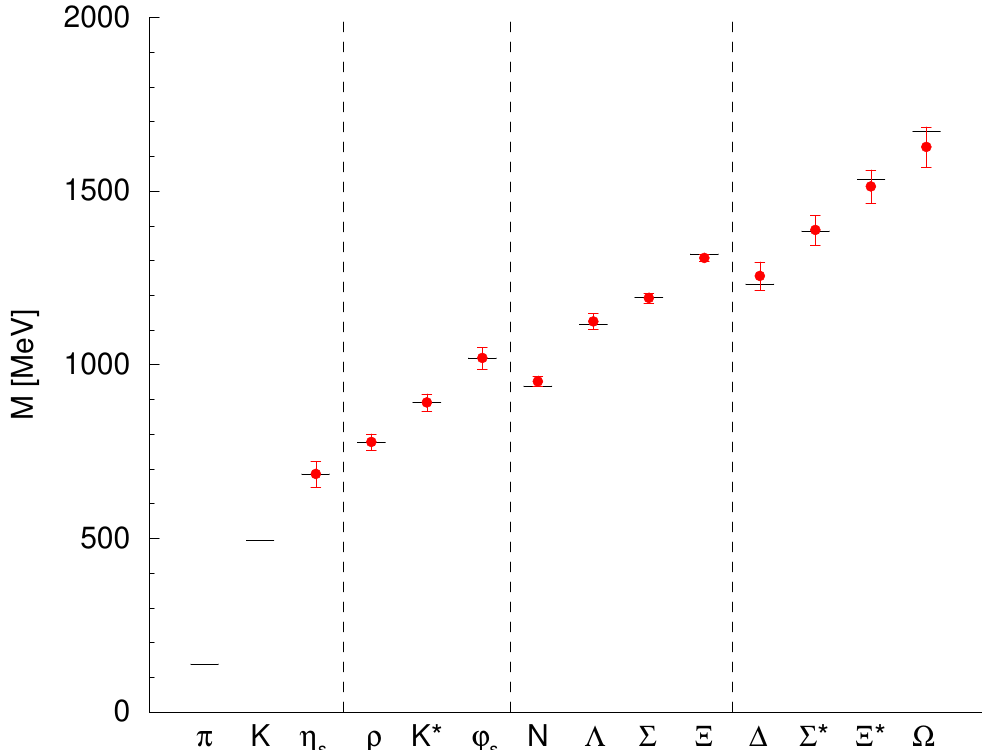}
\vspace{-3mm}
\caption{The light hadron spectrum based on QCD simulations
  (red bullets) compared to the experimental values (black lines)
  \cite{QCD}. They agree to percent-level.
  The simulation uses the pion and kaon mass as an input
  to tune the bare quark masses $m_{u} \simeq m_{d}$ and $m_{s}$.}
\end{center}
\labelf{hadspec}
\vspace{-8.5mm}
\end{figure}

We see that we do have precision-results for lattice QCD now,
despite Wilson's pessimism which he expressed in 1989, before
leaving the field. These are non-perturbative results from first
principles, which demonstrate, for instance, that QCD (rather than
the Higgs mechanism) correctly accounts for almost $99\,\%$ of the mass
of the nucleons, and therefore of all macroscopic objects around us.

In some cases, lattice QCD has even proceeded from post-dictions to
predictions: for instance, the meson mass $M_{B_{c}}$ was predicted by the
HPQCD Collaboration in 2004 as $6.304(12)^{+18}_{0}~{\rm GeV}$, and
experimentally measured a little later by the CDF Collaboration in
excellent agreement, $6.287(5)~{\rm GeV}$, as reviewed in
Ref.\ \cite{Andreas}.

In lattice QCD, work is ongoing on even higher precision
of the hadron masses, matrix elements, excited states,
decay constants etc., with lattice spacings typically in the range
of $0.05 \dots 0.1 ~{\rm fm}$ (even finer lattices lead to problems
due to ``topological freezing''). 

The rest of this article is devoted to the historical evolution
of lattice field theory (which is not limited to lattice QCD).
Much of the results of this work were presented before
at the Lattice Conference 2021 \cite{WBLat21}.

\label{sec:method}
\section{Methodology}

In 1991/2, {\it i.e.}\ just around the time when the ``lattice community''
was fully established\footnote{Its conceptual basis was already on solid
grounds, while computational facilities were rapidly improved and they became
widely accessible.}, the online preprint repository {\tt arXiv} \cite{arXiv}
became operational. Its {\tt hep-lat} section is used comprehensively
by the lattice community, with hardly any exceptions,
which provides a unique opportunity to statistically
monitor the evolution of lattice activities over three decades.

We used {\tt INSPIRE}, the dominant High Energy Physics information
platform, to count the following parameters:
\begin{itemize}
\item $E$: {\bf Entries}, all articles with primary {\tt arXiv} section
{\tt hep-lat}
\item $P$: {\bf Papers}, subset of $E$ published as regular research
  articles
\item $C$: {\bf Citations} to all articles in $E$
\item H\,: {\bf Hirsch Index} \cite{Hirsch}, based on $E$.
\end{itemize}
  
These {\tt hep-lat} data are compared to six related
{\tt arXiv} sections: {\tt hep-ph}, {\tt hep-th}, {\tt gr-qc},
{\tt nucl-th}, {\tt quant-ph}, {\tt cond-mat*} (the star means that
we include all subsections), and to the sum over all 7 sections.

As a single parameter, we are going to refer to the linear combination
\begin{equation}
  \label{Sigmadef}
  \Sigma = E + P + 0.05 C \ .
\end{equation}
These weights are motivated by statistical trends, as we will see below.

We are going to present {\em global} as well as {\em national}
statistics. The latter is divided into {\em extensive} and
{\em intensive} data ({\it i.e.}\ {\em absolute} and {\em relative
to the population} or {\em economy}). For a contextual perspective, we
will also consider two socio-economic indices, which are defined by the
United Nations Development Programme \cite{UNO}:
\begin{itemize}
\item GDP and GDPpp: {\bf Gross Domestic Product}, total and
per person
\item ${\rm EI} = \tfrac{1}{2} ({\rm EYS}/18 + {\rm MYS}/15)$:
{\bf Education Index}, composed of the number of expected schooling
years of children (normalized by the duration for a Master degree)
and the mean schooling years of adults (normalized by the expected
maximum in 2025).\footnote{I also collected data for the
  Human Development Index (HDI), which considers health, income and
  education. However, the values tend to be similar to the EI,
  hence I do not display the HDI. I further took data for the
  percentage of the ``skilled labor force'', but in some cases these
  numbers do not make sense.}
\end{itemize}

In a national statistics, an article counts for a country if at least
one author has a working address there (regardless of the author's
nationality or country of birth). Hence an
article can count for several countries, but in the {\tt hep-lat}
(and to some extent in all 7 {\tt arXiv} sections that we consider)
collaborations tend to be small, which keeps this definition meaningful.

The data were taken in July 2020 from https://old.inspirehep.net;
a little later that INSPIRE version was disactivated, unfortunately.
The new INSPIRE version does not have the search option ``country
code'', which we need for national statistics. The options ``date''
and ``date earliest'' exist, but they return weird results
\cite{WBLat21}, hence temporal data are not accessible anymore either.
\vspace*{-4mm}

\label{sec:global}
\section{Global Statistics}

Figure \ref{evolglob} shows the evolution of the total number of
{\tt hep-lat} entries ($E$), papers ($P$) and citations ($C$) from 1992
to 2019. We exclude 1991 (there were only a few sporadic entries) and
2020 (we could only capture the first half). After an early increase,
$E$ and $P$ have essentially been stagnating since about 1998, at a level
of $\approx 11$ entries a week, and $\approx 5$ of them later turned
into papers.
(The late decrease of $C$ is simply because these articles had little
time to be cited before summer 2020.)
\begin{figure}[h!]
\begin{center}
\vspace{-4mm}
\includegraphics[angle=270,width=95mm]{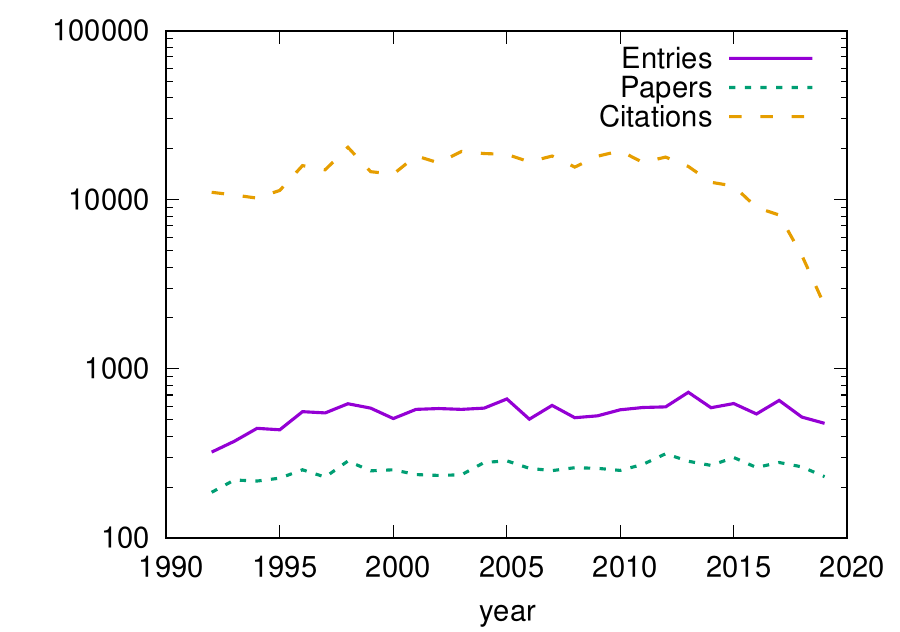}
\vspace{-3mm}
\caption{Time evolution of {\tt hep-lat} entries, papers and
citations, from 1992 to 2019.}
\end{center}
\labelf{evolglob}
\vspace{-5mm}
\end{figure}

The plots in Figure \ref{evolglob7} compare the corresponding
evolution of the 7 {\tt arXiv} sections under consideration. We
see that in other sections, $E$ (solid lines) and $P$ (dashed lines)
have stronger and longer-lasting trends of increase.
\begin{figure}[h!]
\begin{center}
\vspace*{-2mm}    
  \includegraphics[angle=270,width=69mm]{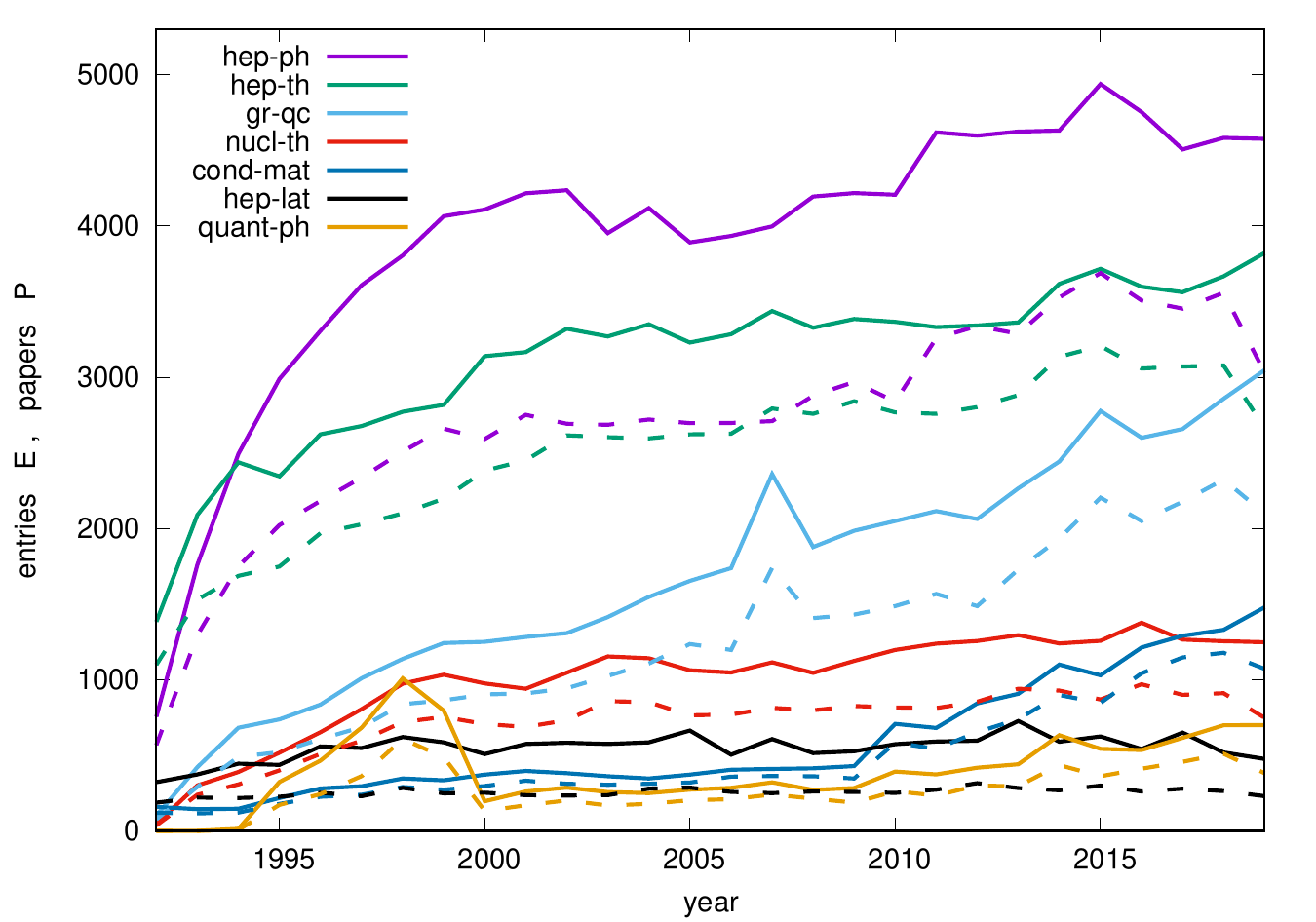}
  \hspace*{-4mm}
  \includegraphics[angle=270,width=69mm]{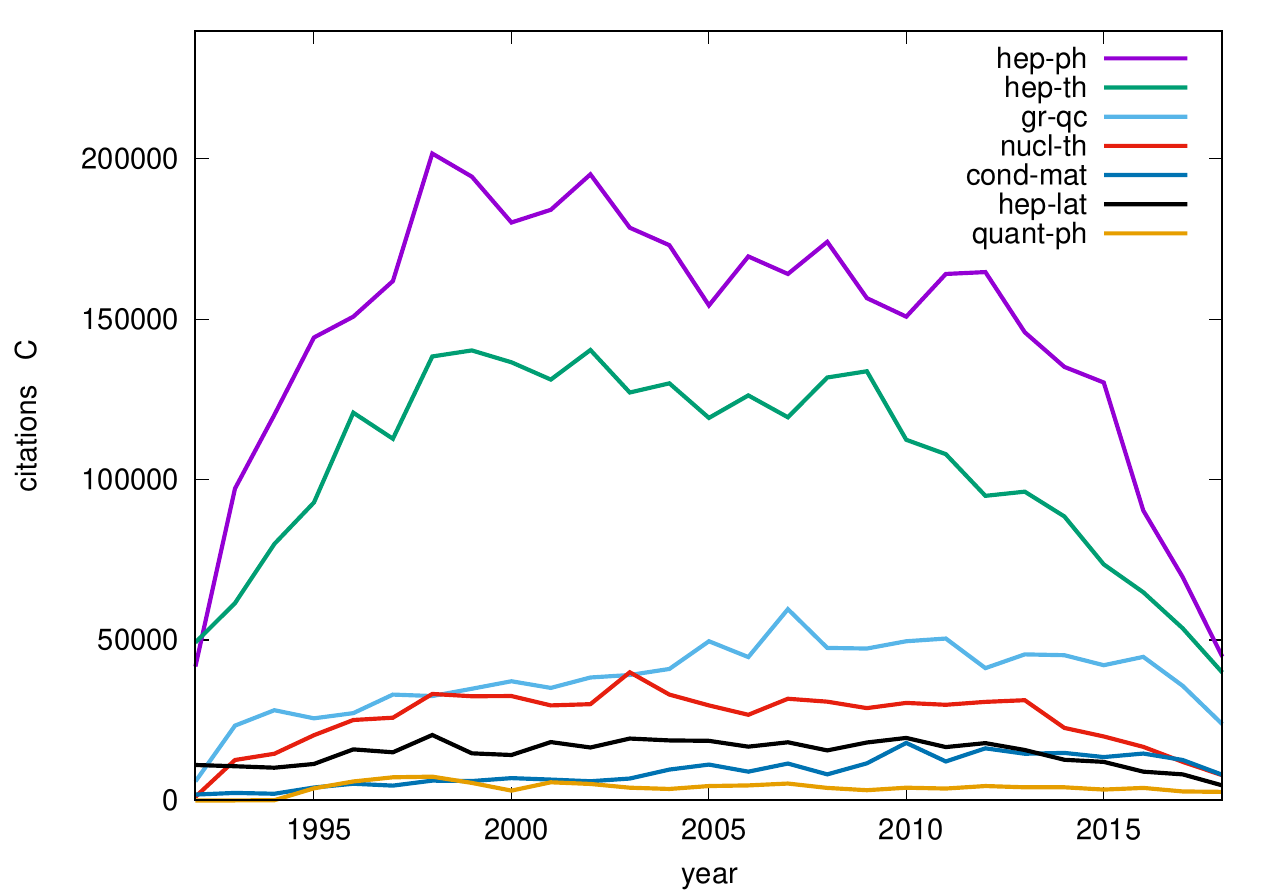}
\vspace{-3mm}
\caption{Time evolution of 7 {\tt arXiv} sections regarding entries
  (left, solid lines), regular papers (left, dashed lines) and citations
  (right).}
\end{center}
\labelf{evolglob7}
\vspace{-6mm}
\end{figure}

Table \ref{tabglob} displays the global statistics for all 7 {\tt arXiv}
sections from 1991 to July 2020. Note that in particular {\tt cond-mat}
captures only part of the extensive work in condensed matter physics,
where it is not a general standard to submit preprints to the {\tt arXiv}.

{\footnotesize
\begin{table}
\centering
\hspace*{-5mm}
\begin{tabular}{|c||r|r|r|r|r|r|r|}
\hline
 & {\tt hep-ph} & {\tt hep-th} & {\tt gr-qc} & {\tt nucl-th}
 & {\tt cond-mat} & {\tt hep-lat} & {\tt quant-ph} \\
\hline
$E$ &  111515 &   89279 &   48927 &  28522 &  16969 &  15610 &  11602 \\
$P$ &   76520 &   70561 &   35703 &  20215 &  13677 &   7165 &   7484 \\
$C$ & 3960720 & 2857462 & 1043823 & 682874 & 247734 & 402121 & 106901 \\
\hline
$P/E$ & 0.69 & 0.79 & 0.71 & 0.71 & 0.81 & 0.46 & 0.65 \\
$C/E$ & 35.5 & 32.0 & 21.3 & 23.9 & 14.6 & 25.8 & 9.21 \\
\hline
\end{tabular}
\caption{Global data for $E$, $P$ and $C$ in 7 {\tt arXiv} sections,
  in the period from 1991 to July 2020.\label{tabglob}}
\vspace*{-2mm}
\end{table}
}

The peculiarity $P < E/2$ of {\tt hep-lat} is most likely
related to the extraordinary role of the proceedings of the Lattice
Conferences: they provide an annual snapshot of the lattice activities,
where almost all participants contribute. The ratio $C/E$ is average
for {\tt hep-lat} (below {\tt hep-ph} and {\tt hep-ph}, similar to
{\tt gr-qc} and {\tt nucl-th}, but above {\tt cond-mat} and {\tt quant-ph}).

\label{sec:extensive}
\section{Extensive National Statistics}

Figure \ref{extentE} shows the evolution of the annual {\tt hep-lat}
entries ($E$) of the 12 leading nations. Initially the USA was dominant,
but since 2010 Germany has caught up, followed by Japan, UK, Italy
and France. We see strong annual peaks in the production of Austria
and Spain, and lately China is moving up (it started with hardly
any lattice activity until 1996).

\begin{figure}[H]
\begin{center}
  \includegraphics[angle=270,width=.505\linewidth]{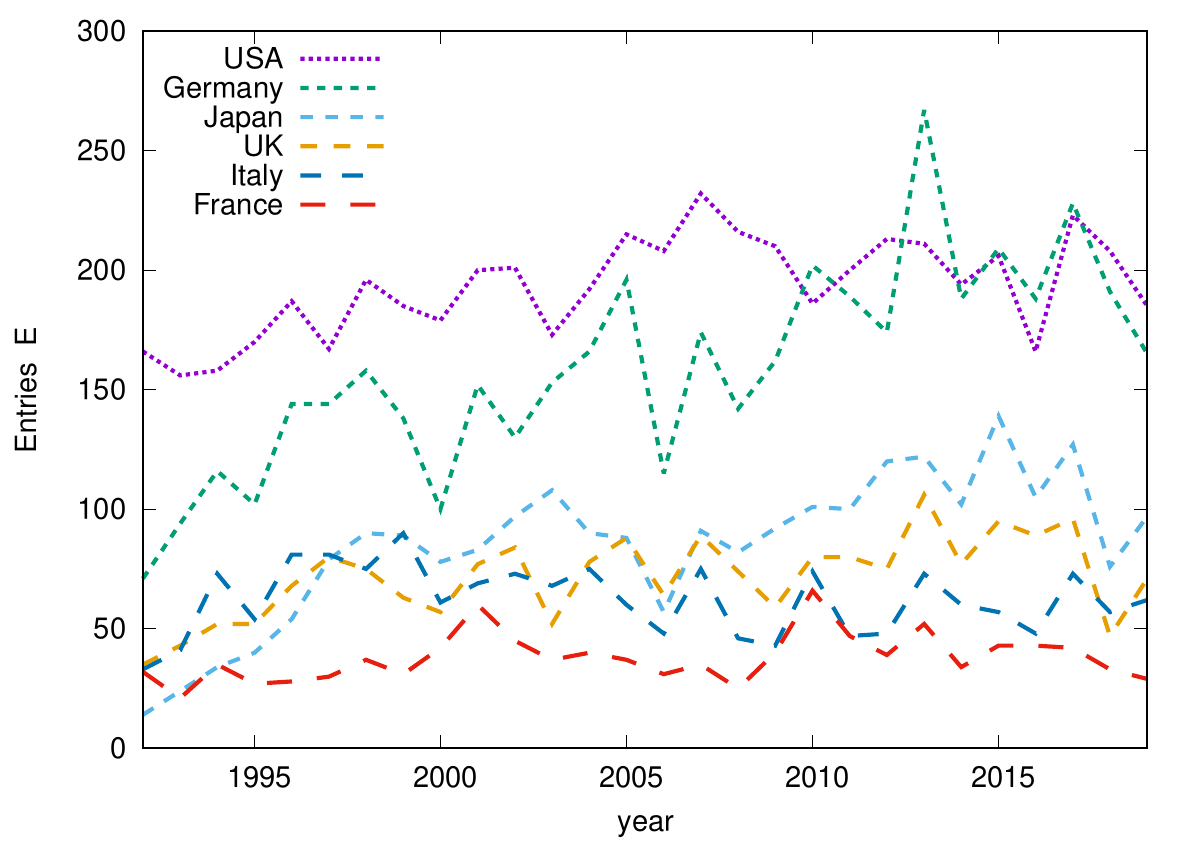}
  \hspace*{-5mm}
  \includegraphics[angle=270,width=.505\linewidth]{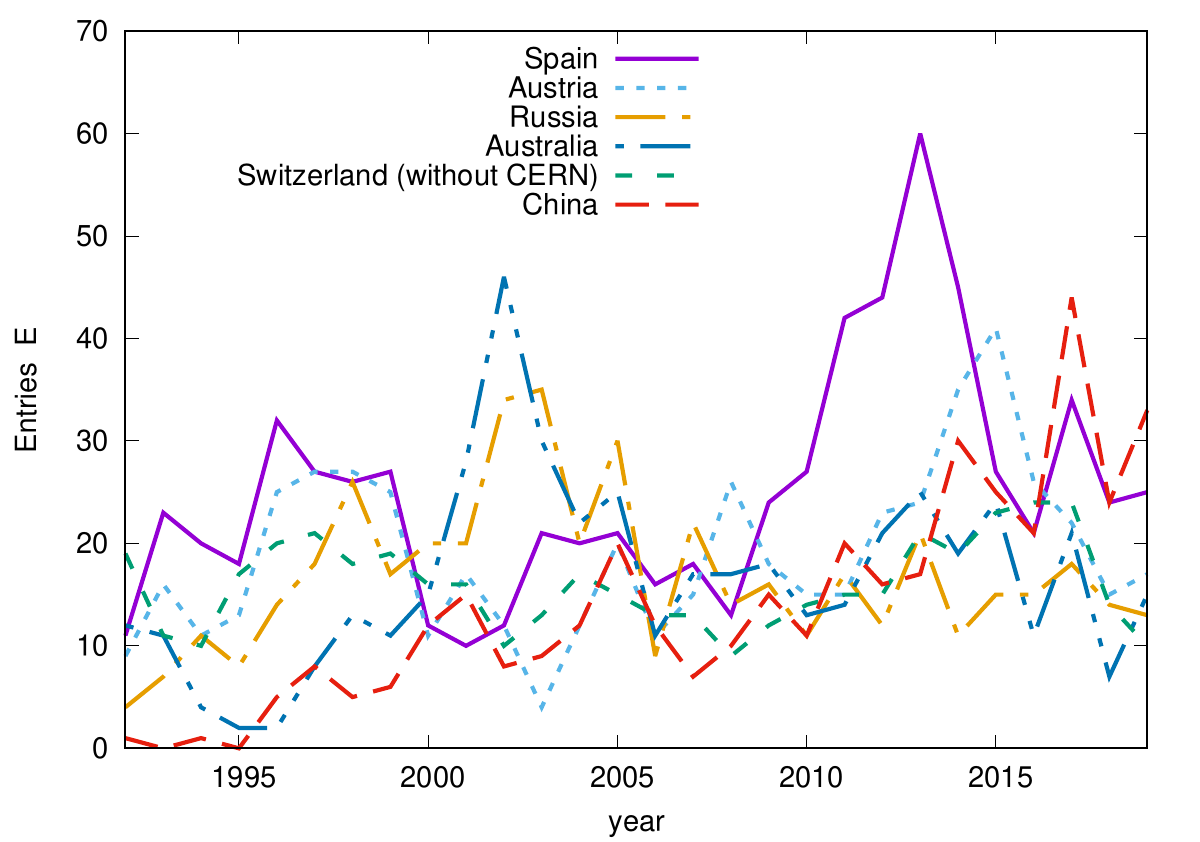} 
\end{center}
\vspace*{-3mm}
\caption{Evolution of the {\tt hep-lat} entries
  of the 6 leading nations (left) and of the nations ranking
  7th to 12th (right), in the period from 1992 to 2019.}
\labelf{extentE}
\vspace*{-1mm}
\end{figure}

Explicit data of the leading nations in terms of the production
in {\tt hep-lat}, and summed over the 7 {\tt arXiv} sections under
consideration, are given in Tables \ref{heplatrankingext} and
\ref{7arxivrankingext}, respectively.
Here the ranking is based on the Hirsch Index H, and it is similar
if we use instead $\Sigma$ from eq.\ (\ref{Sigmadef}).
(In case of identical H index, $\Sigma$ decides about the rank.)
The population is given in millions of inhabitants, and the GDP in
units of $10^{9}$ US \$ (purchasing parity to its value in 2011);
both are averaged over the years from 1992 to 2019.

As benchmarks, we also present data for CERN and for the
European Union, still with 28
nations (including the UK). For a fair comparison, the data for
Switzerland do not include CERN (if no author has another working
address in Switzerland).

\begin{table}
\centering
\begin{footnotesize}
\hspace*{-7mm}
\begin{tabular}{|l c||r|r||r|r|r||c|r|}
  \hline
& $\Sigma$-rank & population & GDP~ & $E$~~ & $P$~~ & $C$~~~ & $\Sigma$ & H \\
\hline
~~~~ European Union
               &    & 495.0 & 15900.4 & 9020 & 4265 & 249236 & {\bf 25746.8} & {\bf 182} \\
~1. USA        &  1 & 294.3 & 14050.1 & 5472 & 2565 & 175639 & {\bf 16818.9} & {\bf 167} \\
~2. Germany    &  2 &  81.3 &  3167.4 & 4523 & 2107 & 143667 & {\bf 13813.3} & {\bf 155} \\
~3. UK         &  3 &  61.4 &  2136.2 & 2026 &  950 &  67386 & {\bf  6345.3} & {\bf 121} \\
~4. Japan      &  4 & 127.6 &  4458.1 & 2419 & 1078 &  55729 & {\bf  6283.5} & {\bf 103} \\
~5. France     &  6 &  61.1 &  2232.9 & 1070 &  540 &  42060 & {\bf  3713.0} & {\bf 101} \\  
~6. Italy      &  5 &  58.5 &  2066.6 & 1765 &  877 &  41252 & {\bf  4704.6} & {\bf  89} \\
~~~~ CERN      &    &       &         &  615 &  306 &  27553 & {\bf  2298.7} & {\bf  79} \\
~7. Switzerland&  7 &   7.5 &   400.7 &  614 &  295 &  24073 & {\bf  2112.7} & {\bf  75} \\
~8. Spain      &  8 &  43.6 &  1326.8 &  708 &  370 &  19310 & {\bf  2043.5} & {\bf  71} \\
~9. Australia  &  9 &  18.9 &   801.9 &  471 &  250 &  14208 & {\bf  1431.4} & {\bf  64} \\
10. Hungary    & 10 &  10.1 &   208.5 &  292 &  151 &  18824 & {\bf  1384.2} & {\bf  61} \\
11. Cyprus     & 14 &   1.0 &    23.1 &  389 &  193 &   8794 & {\bf  1021.7} & {\bf  54} \\  
12. Austria    & 11 &   8.3 &   333.8 &  534 &  254 &  10300 & {\bf  1303.0} & {\bf  51} \\
13. China      & 13 &1326.5 &  9638.8 &  415 &  223 &   9708 & {\bf  1123.4} & {\bf  50} \\
14. Denmark    & 15 &   5.4 &   233.1 &  322 &  174 &  10188 & {\bf  1005.4} & {\bf  50} \\
15. Russia     & 12 & 145.5 &  2822.8 &  475 &  252 &   8608 & {\bf  1157.4} & {\bf  48} \\
16. Canada     & 17 &  32.5 &  1269.4 &  325 &  164 &   8387 & {\bf   908.3} & {\bf  46} \\
17. India      & 16 &1142.0 &  4509.3 &  316 &  183 &   8198 & {\bf   908.9} & {\bf  45} \\
18. Ireland    & 18 &   4.2 &   183.7 &  292 &  117 &   8565 & {\bf   837.3} & {\bf  45} \\
19. Taiwan     & 20 &  22.5 &   702.2 &  196 &  100 &   5491 & {\bf   570.6} & {\bf  40} \\
20. Finland    & 23 &   5.3 &   191.6 &  165 &   81 &   4460 & {\bf   469.0} & {\bf  36} \\
21. Netherlands& 19 &  16.3 &   696.5 &  203 &   98 &   4544 & {\bf   528.2} & {\bf  35} \\ 
22. Israel     & 22 &   6.6 &   197.8 &  149 &   96 &   4622 & {\bf   476.1} & {\bf  35} \\
23. South Korea& 21 &  48.3 &  1242.6 &  238 &   82 &   4448 & {\bf   542.4} & {\bf  34} \\
24. Poland     & 24 &  38.3 &   696.8 &  181 &   82 &   3192 & {\bf   422.6} & {\bf  30} \\
25. Brazil     & 25 & 184.4 &  2397.1 &  117 &   58 &   3939 & {\bf   372.0} & {\bf  29} \\
26. Sweden     & 27 &   9.2 &   365.9 &   99 &   50 &   2162 & {\bf   257.1} & {\bf  27} \\
27. Slovenia   & 30 &   2.0 &    51.1 &   67 &   33 &   2174 & {\bf   208.7} & {\bf  26} \\ 
28. Portugal   & 26 &  10.4 &   264.0 &  138 &   55 &   2252 & {\bf   305.6} & {\bf  25} \\
29. Slovakia   & 28 &   5.4 &   111.5 &   82 &   42 &   2354 & {\bf   241.7} & {\bf  25} \\
30. Greece     & 29 &  10.9 &   273.9 &   96 &   49 &   1756 & {\bf   232.8} & {\bf  22} \\
31. Belgium    & 31 &  10.7 &   409.5 &   54 &   28 &   1792 & {\bf   171.6} & {\bf  20} \\   
32. Mexico     & 33 & 106.6 &  1710.8 &   58 &   26 &    760 & {\bf   122.0} & {\bf  15} \\
33. Ukraine    & 32 &  47.5 &   319.9 &   85 &   34 &    417 & {\bf   139.9} & {\bf  11} \\
34. Turkey     & 34 &  68.2 &  1192.0 &   25 &   19 &    347 & {\bf    61.3} & {\bf  10} \\
35. New Zealand& 36 &   4.1 &   127.7 &   11 &    8 &    420 & {\bf    40.0} & {\bf 8} \\ 
36. Bangladesh & 39 & 137.0 &   317.0 &   10 &    6 &    267 & {\bf   29.4} & {\bf 7} \\
37. Belarus    & 35 &   9.7 &   113.6 &    9 &    6 &    527 & {\bf   41.3} & {\bf 6} \\
38. Iran       & 37 &  69.9 &  1088.1 &   17 &   11 &    126 & {\bf   34.3} & {\bf 6} \\
39. Georgia    & 38 &   4.4 &    22.6 &   11 &    7 &    236 & {\bf   29.8} & {\bf 5} \\
40. Norway     & 40 &   4.7 &   280.4 &   14 &    6 &     83 & {\bf   24.2} & {\bf 5} \\
41. Albania     & 41 &   3.0 &    22.9 &   15 &    2 &    107 & {\bf 22.4} & {\bf 5} \\
42. Singapore   & 43 &   4.5 &   297.6 &   10 &    5 &    113 & {\bf 20.7} & {\bf 4} \\
43. Croatia     & 44 &   4.4 &    78.6 &    9 &    4 &     61 & {\bf 16.1} & {\bf 4} \\
44. Uruguay     & 45 &   3.3 &    50.2 &    7 &    5 &     27 & {\bf 13.4} & {\bf 4} \\
45. Czech Rep.  & 46 &  10.4 &   261.4 &    5 &    5 &     66 & {\bf 13.3} & {\bf 4} \\
46. Chile       & 47 &  16.2 &   281.0 &    7 &    3 &     34 & {\bf 11.7} & {\bf 3} \\  
47. Algeria     & 52 &  33.8 &   404.7 &    3 &    2 &     55 & {\bf  7.8} & {\bf 3} \\   
48. Thailand    & 42 &  64.7 &   774.0 &   10 &    8 &     65 & {\bf 21.3} & {\bf 2} \\  
49. Pakistan    & 48 & 161.4 &   635.0 &    7 &    2 &     14 & {\bf  9.7} & {\bf 2} \\   
50. Armenia     & 49 &   3.0 &    15.5 &    7 &    2 &      5 & {\bf  9.3} & {\bf 2} \\ 
\hline
\end{tabular}
\end{footnotesize}
\caption{Extensive {\tt hep-lat} statistics for the 50 leading nations,
  ranked by the Hirsch Index H. Based on the $\Sigma$-rank, South Africa
  enters the top 50, at position~50.}
\label{heplatrankingext}
\end{table}

\begin{table}
\centering
\begin{footnotesize} 
\begin{tabular}{|l c||r|r|r||r|r|}
  \hline
& $\Sigma$-rank & $E$~~ & $P$~~ & $C$~~~ & $\Sigma$~~~~~ & H \\
\hline
~1. USA         &  1 &  85601 &  63422 & 4006414 & {\bf 349343.7} & {\bf 583} \\
~~~~ European
Union           &    & 147464 & 106804 & 4961069 & {\bf 502321.4} & {\bf 548} \\
~2. France      &  3 &  28598 &  21090 & 1368170 & {\bf 118096.5} & {\bf 394} \\ 
~3. Germany     &  2 &  45242 &  35533 & 1713406 & {\bf 166445.3} & {\bf 390} \\
~4. UK          &  4 &  26174 &  19927 & 1089933 & {\bf 100597.7} & {\bf 339} \\
~~~~ CERN       &    &   9836 &   7253 &  682730 & {\bf  51225.5} & {\bf 333} \\
~5. Italy       &  5 &  27034 &  19875 &  949801 & {\bf  94399.1} & {\bf 309} \\
~6. Spain       &  9 &  16466 &  12223 &  601111 & {\bf  58744.6} & {\bf 267} \\
~7. Russia      &  7 &  22429 &  15260 &  586236 & {\bf  67000.8} & {\bf 257} \\
~8. Canada      & 10 &  12364 &   9696 &  469849 & {\bf  45552.5} & {\bf 250} \\
~9. Japan       &  6 &  24611 &  18620 &  690235 & {\bf  77742.8} & {\bf 249} \\
10. Switzerland & 14 &  5734  &   4142 &  290056 & {\bf  24378.8} & {\bf 227} \\
11. Netherlands & 16 &   5634 &   4307 &  262786 & {\bf  23080.3} & {\bf 207} \\
12. Poland      & 13 &   8205 &   5660 &  245108 & {\bf  26120.4} & {\bf 191} \\
13. China       &  8 &  20475 &  16372 &  444642 & {\bf  59079.1} & {\bf 186} \\
14. India       & 11 &  14374 &  10925 &  332859 & {\bf  41942.0} & {\bf 186} \\
15. Sweden      & 18 &   4973 &   3753 &  198321 & {\bf  18642.1} & {\bf 178} \\
16. Belgium     & 17 &   4955 &   3798 &  201122 & {\bf  18809.1} & {\bf 171} \\
17. Israel      & 19 &   5085 &   3943 &  176081 & {\bf  17832.1} & {\bf 166} \\
18. Brazil      & 12 &  11460 &   8961 &  222670 & {\bf  31554.5} & {\bf 146} \\
19. South Korea & 15 &   7627 &   6096 &  196954 & {\bf  23570.7} & {\bf 142} \\
20. Portugal    & 22 &   4166 &   3087 &  128890 & {\bf  13697.5} & {\bf 139} \\
21. Austria     & 23 &   3885 &   2563 &  112541 & {\bf  12075.1} & {\bf 137} \\
22. Greece      & 24 &   3505 &   2748 &  107266 & {\bf  11616.3} & {\bf 134} \\
23. Australia   & 20 &   4185 &   3219 &  136125 & {\bf  14210.3} & {\bf 133} \\
24. Denmark     & 27 &   3006 &   2248 &  103328 & {\bf  10420.4} & {\bf 133} \\
25. Taiwan      & 21 &   4110 &   3186 &  134243 & {\bf  14008.2} & {\bf 129} \\
26. Finland     & 29 &   2460 &   1818 &   86101 & {\bf   8583.1} & {\bf 128} \\
27. Hungary     & 30 &   2274 &   1586 &   88347 & {\bf   8277.4} & {\bf 121} \\
28. Chile       & 28 &   3089 &   2567 &   77183 & {\bf   9515.2} & {\bf 108} \\
29. Argentina   & 31 &   2576 &   2179 &   63899 & {\bf   7950.0} & {\bf 104} \\
30. Mexico      & 25 &   4450 &   3232 &   75687 & {\bf  11466.4} & {\bf 101} \\
31. Ireland     & 36 &   1476 &   1078 &   43298 & {\bf   4718.9} & {\bf  97} \\
32. Iran        & 26 &   4054 &   3398 &   64526 & {\bf  10678.3} & {\bf  89} \\
33. South Africa & 34 &   1903 &   1483 &   44556 & {\bf 5613.8} & {\bf 85} \\
34. Slovenia     & 41 &    718 &    484 &   28219 & {\bf 2613.0} & {\bf 85} \\
35. Norway       & 37 &   1242 &    923 &   35198 & {\bf 3924.9} & {\bf 84} \\
36. Ukraine      & 33 &   2508 &   1674 &   42411 & {\bf 6302.6} & {\bf 82} \\
37. Czech Rep.   & 35 &   2106 &   1501 &   37754 & {\bf 5494.7} & {\bf 78} \\
38. Bulgaria     & 40 &   1040 &    684 &  24849  & {\bf 2966.5} & {\bf 73} \\
39. Croatia      & 38 &   1035 &    778 &   25216 & {\bf 3073.8} & {\bf 71} \\
40. Estonia      & 47 &    425 &    336 &   20056 & {\bf 1763.8} & {\bf 71} \\
41. Turkey       & 32 &   2571 &   2112 &   32563 & {\bf 6311.2} & {\bf 62} \\
42. Georgia      & 45 &    693 &    512 &   16033 & {\bf 2006.7} & {\bf 62} \\
43. Armenia      & 42 &    835 &    635 &   17331 & {\bf 2336.6} & {\bf 61} \\
44. Slovakia     & 48 &    642 &    394 &   13173 & {\bf 1694.7} & {\bf 61} \\
45. New Zealand  & 49 &    495 &    362 &   14576 & {\bf 1585.8} & {\bf 60} \\
46. Cyprus       & 51 &    497 &    271 &   11675 & {\bf 1351.8} & {\bf 59} \\
47. Venezuela    & 50 &    489 &    394 &   11564 & {\bf 1461.2} & {\bf 57} \\
48. Colombia     & 43 &    842 &    623 &   15975 & {\bf 2263.8} & {\bf 56} \\
49. Romania      & 39 &   1256 &    913 &   17598 & {\bf 3048.9} & {\bf 55} \\
50. Egypt        & 46 &    738 &    559 &   12755 & {\bf 1934.8} & {\bf 53} \\
\hline
\end{tabular}
\end{footnotesize}
\caption{Like Table \ref{heplatrankingext}, but joint
statistics for the 7 {\tt arXiv} sections.
Based on the $\Sigma$-rank, Pakistan enters the top 50,
at position 44.}
\label{7arxivrankingext}
\end{table}

\label{sec:intensive}
\section{Intensive National Statistics}

\begin{figure}[h!]
\begin{center}
\includegraphics[angle=0,width=0.495\linewidth]{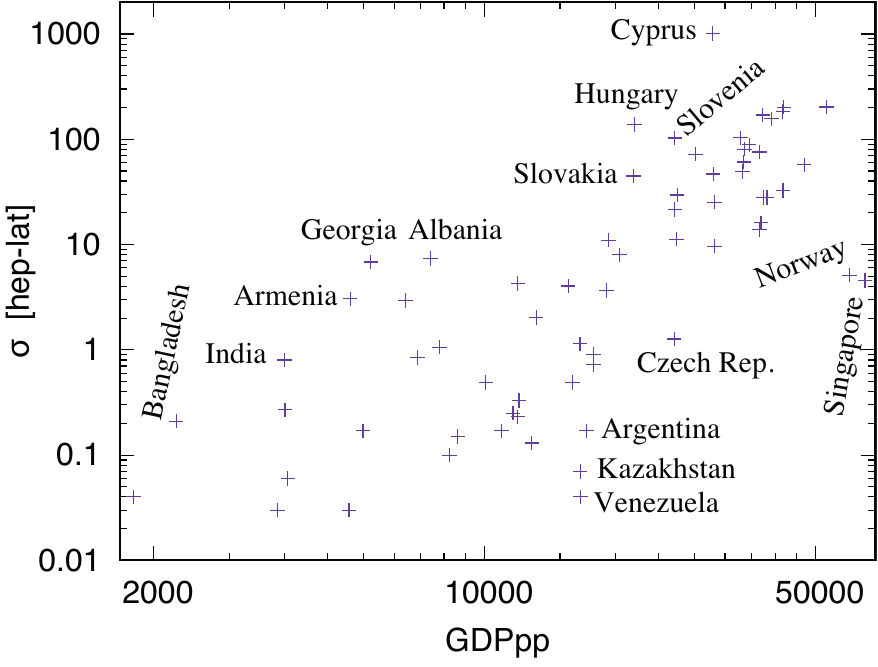}
\hspace*{-2mm}
\includegraphics[angle=0,width=0.495\linewidth]{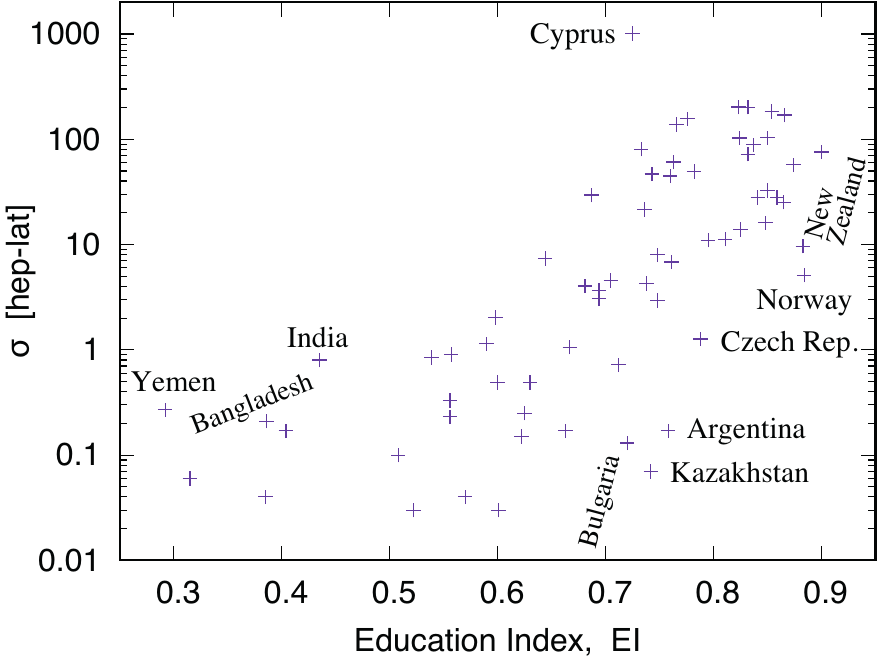}
\end{center}
\vspace*{-6mm}
\caption{Scatter plots for $\sigma$ (defined in eq.\ (\ref{sigmadef}),
  based on {\tt hep-lat} data) against the GDPpp (left) and against
  the EI (right).}
\label{scatheplat}
\vspace*{-1mm}
\end{figure}

In total, there are 66 nations with {\tt hep-lat} entries. Figure
\ref{scatheplat} shows two scatter plots for all of them, with
intrinsic quantities. We normalize the term $\Sigma$ of eq.\
(\ref{Sigmadef}) by the population,
\begin{equation}
  \label{sigmadef}
  \sigma := \frac{\Sigma}{\rm population~in~millions} \ ,
\end{equation}
and plot it against the GDPpp (left) and the EI (right).
A monotonic trend is visible, but not as clearly as one might
have expected --- the scattering is rather broad. Nations with a
particularly high or low {\tt hep-lat} production per capita, with
respect to their economic or educational potential, are indicated in
the plots. The top 50 nations can be identified from
Table \ref{heplatrankingint}.

\begin{figure}[h!]
\begin{center}
\includegraphics[angle=0,width=.495\linewidth]{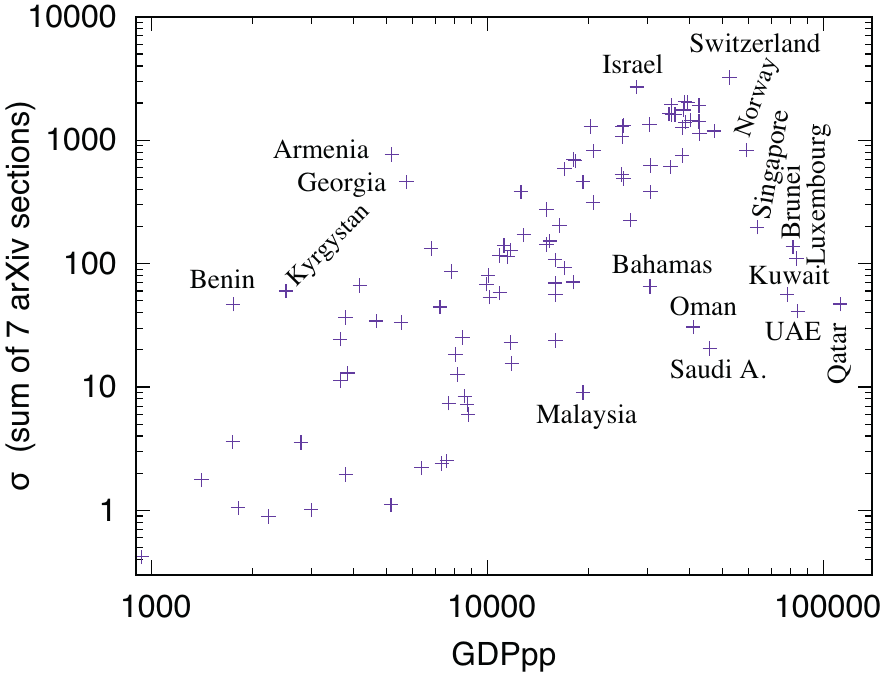}
\hspace*{-2mm}
\includegraphics[angle=0,width=.495\linewidth]{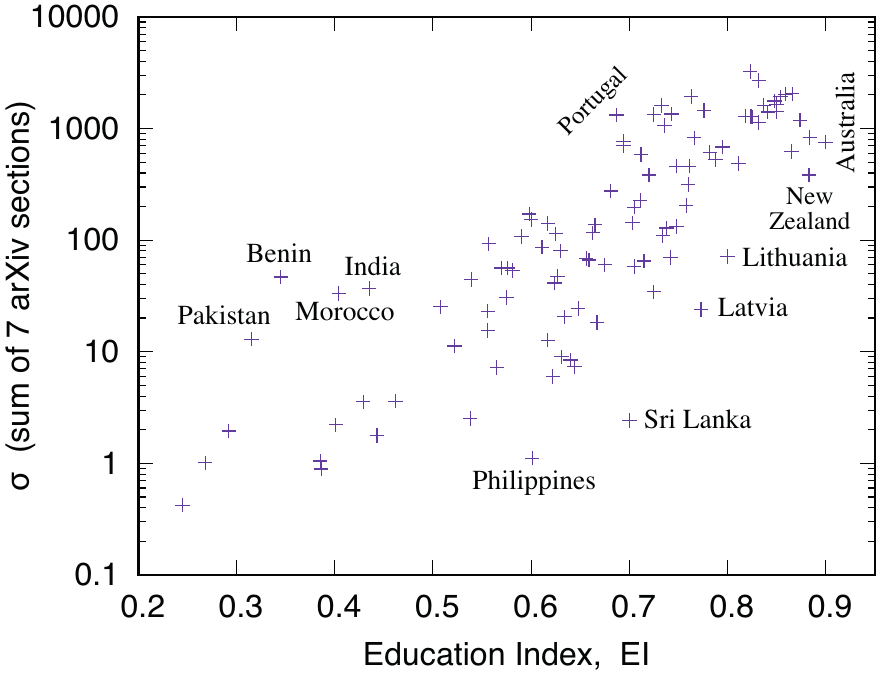}
\end{center}
\vspace*{-6mm}
\caption{Scatter plots for $\sigma$ based on the data summed over the
  7 {\tt arXiv} sections, against the GDPpp (left)
  and against the EI (right).}
\label{scat7arxiv}
\vspace*{-2mm}
\end{figure}

The analogous plots for the 7 {\tt arXiv} sections are shown in
Figure \ref{scat7arxiv}. Here we include the 100 nations with
${\rm H} \geq 5$, and the leading 50 nations can be identified
from Table \ref{7arxivrankingint}. The monotonic trend is somewhat
clearer than in Figure \ref{scatheplat}, which is more specific.
Again the countries which deviate most from this trend are indicated
inside the plots.

As we anticipated, Tables \ref{heplatrankingint} and \ref{7arxivrankingint}
are devoted to intrinsic data, based on {\tt hep-lat} (Table
\ref{heplatrankingint}) and the 7 {\tt arXiv} sections (Table
\ref{7arxivrankingint}, were we exclude tiny nations with less than
$50\,000$ inhabitants, in particular Monaco and the Vatican).
Here the ranking refers to the parameter $\sigma$.
We further include the normalized --- and therefore intrinsic --- versions
of quantities introduced in Section 2,
\begin{equation} \label{epcnorm}
(e,p,c) := \frac{(E,P,C)}{\rm population~in~millions} \ ,
\end{equation}
along with the GDPpp and the EI. We also display the economic rank (e-rank),
which refers to $\Sigma/{\rm GDP} \propto \sigma / {\rm GDPpp}$.

\begin{table}
\label{heplatrankingint}
\vspace*{-2mm}
\centering
\begin{footnotesize} 
\hspace*{-11mm}
\begin{tabular}{|l c||r|r||r|r|r||r|r|}
  \hline
& e-rank & GDPpp & EI~ & $e$~~~ & $p$~~~ & $c$~~~ & $\Sigma$/GDP & $\sigma$~~~~~ \\
\hline
~1. Cyprus      &  1 & 30386 & 0.725 & 383.32 & 190.18 & 8665.6 & {\bf 44.17} & {\bf 1006.78} \\
~2. Switzerland &  3 & 52698 & 0.823 &  81.46 &  39.14 & 3194.0 & {\bf  5.27} & {\bf  280.30} \\
~3. Ireland     &  4 & 42903 & 0.832 &  70.27 &  28.16 & 2061.1 & {\bf  4.56} & {\bf  201.48} \\
~4. Denmark     &  6 & 42671 & 0.854 &  59.22 &  32.00 & 1873.8 & {\bf  4.31} & {\bf  184.02} \\
~5. Germany     &  5 & 38719 & 0.866 &  55.60 &  25.90 & 1766.1 & {\bf  4.36} & {\bf  169.81} \\
~6. Austria     &  8 & 40348 & 0.776 &  64.31 &  30.59 & 1240.4 & {\bf  3.90} & {\bf  156.92} \\
~7. Hungary     &  2 & 20738 & 0.766 &  29.04 &  15.02 & 1872.0 & {\bf  6.64} & {\bf  137.66} \\
~8. UK          &  9 & 34770 & 0.850 &  32.99 &  15.47 & 1097.4 & {\bf  2.97} & {\bf  103.33} \\
~9. Slovenia    &  7 & 25235 & 0.824 &  33.01 &  16.26 & 1071.1 & {\bf  4.08} & {\bf  102.83} \\
10. Finland     & 10 & 36205 & 0.837 &  31.18 &  15.30 &  842.7 & {\bf  2.45} & {\bf   88.61} \\
11. Italy       & 12 & 35411 & 0.733 &  30.19 &  15.00 &  705.6 & {\bf  2.28} & {\bf   80.47} \\
12. Australia   & 14 & 38067 & 0.900 &  24.95 &  13.25 &  752.8 & {\bf  1.79} & {\bf   75.84} \\
13. Israel      & 11 & 27906 & 0.832 &  22.58 &  14.55 &  700.3 & {\bf  2.41} & {\bf   72.14} \\
14. France      & 15 & 35303 & 0.763 &  17.52 &   8.42 &  688.8 & {\bf  1.66} & {\bf   60.80} \\
15. USA         & 19 & 47324 & 0.874 &  18.59 &   8.72 &  596.9 & {\bf  1.20} & {\bf   57.15} \\
~~~~ European
Union           &    & 31966 & 0.792 &  18.22 &   8.62 &  503.5 & {\bf  1.62} & {\bf   52.01} \\
16. Japan       & 17 & 35125 & 0.782 &  18.96 &   8.45 &  436.8 & {\bf  1.41} & {\bf   49.25} \\
17. Spain       & 16 & 30406 & 0.743 &  16.24 &   8.48 &  442.8 & {\bf  1.54} & {\bf   46.86} \\
18. Slovakia    & 13 & 20657 & 0.760 &  15.19 &   7.78 &  435.9 & {\bf  2.17} & {\bf   44.76} \\
19. Netherlands & 24 & 42689 & 0.850 &  12.48 &   6.02 &  279.3 & {\bf  0.76} & {\bf   32.46} \\
20. Portugal    & 20 & 25515 & 0.687 &  13.33 &   5.31 &  217.5 & {\bf  1.16} & {\bf   29.52} \\
21. Canada      & 25 & 38828 & 0.841 &  10.01 &   5.05 &  258.4 & {\bf  0.72} & {\bf   27.98} \\
22. Sweden      & 26 & 39526 & 0.859 &  10.77 &   5.44 &  235.3 & {\bf  0.70} & {\bf   27.98} \\
23. Taiwan      & 23 & 30654 & 0.865 &   8.70 &   4.44 &  243.7 & {\bf  0.81} & {\bf   25.33} \\
24. Greece      & 22 & 25240 & 0.736 &   8.84 &   4.51 &  161.7 & {\bf  0.85} & {\bf   21.43} \\
25. Belgium     & 31 & 38398 & 0.848 &   5.07 &   2.63 &  168.1 & {\bf  0.42} & {\bf   16.10} \\
26. Iceland     & 34 & 38087 & 0.825 &   3.33 &   3.33 &  143.3 & {\bf  0.36} & {\bf   13.83} \\
27. South Korea & 30 & 25473 & 0.811 &   4.93 &   1.70 &   92.1 & {\bf  0.44} & {\bf   11.24} \\
28. Poland      & 27 & 18267 & 0.795 &   4.72 &   2.14 &   83.3 & {\bf  0.61} & {\bf   11.02} \\
29. New Zealand & 35 & 30601 & 0.883 &   2.66 &   1.93 &  101.4 & {\bf  0.31} & {\bf    9.66} \\
30. Russia      & 32 & 19274 & 0.748 &   3.27 &   1.73 &   59.2 & {\bf  0.41} & {\bf    7.96} \\
31. Albania     & 21 &  7683 & 0.644 &   4.96 &   0.66 &   35.4 & {\bf  0.98} & {\bf    7.39} \\
32. Georgia     & 18 &  5752 & 0.761 &   2.52 &   1.60 &   54.0 & {\bf  1.32} & {\bf    6.82} \\
33. Norway      & 43 & 59066 & 0.884 &   2.96 &   1.27 &   17.6 & {\bf  0.09} & {\bf    5.11} \\
34. Singapore   & 46 & 63641 & 0.705 &   2.22 &   1.11 &   25.0 & {\bf 0.07} & {\bf 4.58} \\
35. Belarus     & 33 & 11763 & 0.738 &   0.93 &   0.62 &   54.4 & {\bf 0.36} & {\bf 4.27} \\
36. Uruguay     & 36 & 15032 & 0.681 &   2.11 &   1.51 &    8.1 & {\bf 0.27} & {\bf 4.03} \\
37. Croatia     & 37 & 18091 & 0.694 &   2.05 &   0.91 &   13.9 & {\bf  0.20} & {\bf    3.65} \\ 
38. Armenia     & 28 &  5207 & 0.694 &   2.30 &   0.66 &    1.6 & {\bf  0.60} & {\bf    3.05} \\   
39. Ukraine     & 29 &  6806 & 0.748 &   1.79 &   0.72 &    8.8 & {\bf  0.44} & {\bf    2.95} \\
40. Brazil      & 39 & 12856 & 0.598 &   0.63 &   0.31 &   21.4 & {\bf  0.16} & {\bf    2.02} \\
41. Czech Rep.  & 48 & 25149 & 0.788 &   0.48 &   0.48 &    6.3 & {\bf  0.05} & {\bf    1.27} \\
42. Mexico      & 45 & 15910 & 0.590 &   0.54 &   0.24 &    7.1 & {\bf  0.07} & {\bf    1.14} \\
43. Jordan      & 40 &  8043 & 0.667 &   0.46 &   0.46 &    2.6 & {\bf  0.13} & {\bf    1.06} \\ 
44. Turkey      & 47 & 16994 & 0.557 &   0.37 &   0.28 &    5.1 & {\bf  0.05} & {\bf    0.90} \\
45. China       & 41 &  7231 & 0.539 &   0.31 &   0.17 &    7.3 & {\bf  0.12} & {\bf    0.85} \\
46. India       & 38 &  3782 & 0.435 &   0.28 &   0.16 &    7.2 & {\bf 0.20}  & {\bf    0.80} \\
47. Chile       & 50 & 16993 & 0.712 &   0.43 &   0.19 &    2.1 & {\bf 0.04}  & {\bf    0.72} \\
48. Iran        & 51 & 15355 & 0.600 &   0.24 &   0.16 &    1.8 & {\bf 0.03}  & {\bf    0.49} \\
49. N.\ Macedonia & 49 & 10066 & 0.630 & 0.49 &   0.00 &    0.0 & {\bf 0.05}  & {\bf    0.49} \\
50. Thailand    & 53 & 11830 & 0.556 &   0.15 &   0.12 &    1.0 & {\bf 0.03}  & {\bf    0.33} \\
\hline
\end{tabular}
\end{footnotesize}
\vspace*{-2mm}
\caption{Intensive {\tt hep-lat} statistics,
  ordered according to the parameter $\sigma$
  of eqs.\ (\ref{Sigmadef}) and (\ref{sigmadef}).
  We include further intensive quantities: GDPpp, EI and $(e,p,c)$
  of eq.\ (\ref{epcnorm}). We also display the e-rank, based on
  $\Sigma/{\rm GDP} \propto \sigma/{\rm GPDpp}$.
  With that respect, Bangladesh enters the top 50,
  at position 42.}
\label{heplatrankingint}
\end{table}

\begin{table}
\vspace*{-3mm}
\centering
\begin{footnotesize}
\begin{tabular}{|l c||r|r|r||r|r|}
  \hline
& e-rank & $e$~~~ & $p$~~~ & $c$~~~ & $\Sigma$/GDP & $\sigma$~~~~~ \\
\hline
1.~~ Switzerland       &  5 & 760.78 & 549.55 & 38484.1 & {\bf  60.84} & {\bf 3234.53} \\
2.~~ Israel            &  2 & 770.45 & 597.42 & 26678.9 & {\bf  90.16} & {\bf 2701.83} \\ 
3.~~ Germany           &  8 & 556.15 & 436.80 & 21062.6 & {\bf  52.55} & {\bf 2046.09} \\
4.~~ Sweden            & 11 & 541.20 & 408.43 & 21582.7 & {\bf  50.95} & {\bf 2028.76} \\
5.~~ France            &  7 & 468.31 & 345.36 & 22404.5 & {\bf  52.89} & {\bf 1933.89} \\
6.~~ Denmark           & 16 & 552.87 & 413.46 & 19004.5 & {\bf  44.70} & {\bf 1916.56} \\  
7.~~ Belgium           & 13 & 464.85 & 356.31 & 18868.3 & {\bf  45.93} & {\bf 1764.58} \\
8.~~ United Kingdom    & 12 & 426.24 & 324.50 & 17749.2 & {\bf  47.09} & {\bf 1638.20} \\
9.~~ Finland           & 15 & 464.80 & 343.50 & 16268.2 & {\bf  44.79} & {\bf 1621.71} \\
10.~ Italy             & 14 & 462.38 & 399.94 & 16245.2 & {\bf  45.68} & {\bf 1614.58} \\
11.~ Austria           & 27 & 467.86 & 308.66 & 13553.1 & {\bf  30.09} & {\bf 1454.18} \\
12.~ Netherlands       & 24 & 346.27 & 264.71 & 16151.2 & {\bf  33.14} & {\bf 1418.55} \\
13.~ Canada            & 22 & 380.91 & 298.71 & 14475.0 & {\bf  35.89} & {\bf 1403.37} \\
14.~ Spain             & 17 & 377.60 & 280.30 & 13784.6 & {\bf  44.28} & {\bf 1347.12} \\
15.~ Cyprus            &  6 & 489.74 & 267.04 & 11504.6 & {\bf  58.43} & {\bf 1332.02} \\
16.~ Portugal          &  9 & 402.44 & 298.21 & 12450.9 & {\bf  51.88} & {\bf 1323.19} \\
17.~ Estonia           &  4 & 310.98 & 245.85 & 14675.1 & {\bf  64.03} & {\bf 1290.59} \\
18.~ Slovenia          & 10 & 353.76 & 238.47 & 13903.5 & {\bf  51.14} & {\bf 1287.40} \\
19.~ Iceland           & 25 & 476.67 & 380.00 & 8336.7  & {\bf  33.02} & {\bf 1273.50} \\
20.~ United States     & 30 & 290.89 & 215.52 & 13614.6 & {\bf  24.86} & {\bf 1187.14} \\
21.~ Ireland           & 29 & 355.19 & 259.41 & 10419.3 & {\bf  25.69} & {\bf 1135.56} \\
22.~ Greece            & 18 & 322.66 & 252.97 &  9874.5 & {\bf  42.41} & {\bf 1069.35} \\
~~~~~ European Union   &    & 297.91 & 215.77 & 10022.5 & {\bf  31.59} & {\bf 1014.81} \\
23.~ Norway            & 42 & 262.81 & 195.31 &  7447.9 & {\bf  14.00} & {\bf  830.50} \\
24.~ Hungary           & 19 & 226.14 & 157.72 &  8785.9 & {\bf  39.70} & {\bf  823.16} \\
25.~ Armenia           &  1 & 274.94 & 209.09 &  5706.5 & {\bf 150.46} & {\bf  769.35} \\ 
26.~ Australia         & 39 & 221.73 & 170.55 &  7212.3 & {\bf  17.72} & {\bf  752.90} \\
27.~ Croatia           & 20 & 235.43 & 176.97 &  5735.7 & {\bf  39.12} & {\bf  699.18} \\
28.~ Poland            & 21 & 214.02 & 147.64 &  6393.5 & {\bf  37.49} & {\bf  681.34} \\
29.~ Taiwan            & 34 & 182.44 & 141.43 &  5959.0 & {\bf  19.95} & {\bf  621.82} \\
30.~ Japan             & 40 & 192.89 & 145.94 &  5409.8 & {\bf  17.44} & {\bf  609.32} \\
31.~ Chile             & 23 & 190.64 & 158.42 &  4763.3 & {\bf 33.86} & {\bf  587.22} \\  
32.~ Czech Rep.        & 33 & 201.85 & 143.87 &  3618.6 & {\bf 21.02} & {\bf  526.65} \\
33.~ South Korea       & 36 & 158.01 & 126.29 &  4080.2 & {\bf 18.97} & {\bf  488.31} \\
34.~ Russia            & 31 & 154.18 & 104.90 &  4029.9 & {\bf 23.74} & {\bf  460.58} \\
35.~ Georgia           &  3 & 158.70 & 117.25 &  3671.7 & {\bf 88.78} & {\bf  459.54} \\
36.~ New Zealand       & 46 & 119.54 &  87.42 &  3520.1 & {\bf 12.42}  & {\bf  382.98} \\
37.~ Bulgaria          & 26 & 134.16 &  88.24 &  3205.6 & {\bf 30.90}  & {\bf  382.68} \\
38.~ Slovakia          & 41 & 118.89 &  72.96 &  2439.4 & {\bf 15.20}  & {\bf  313.82} \\
39.~ Uruguay           & 38 &  92.61 &  76.32 &  2119.0 & {\bf 18.14}  & {\bf  274.89} \\
40.~ Malta             & 54 &  82.50 &  72.50 &  1405.0 & {\bf  8.18}  & {\bf  225.25} \\
41.~ Argentina         & 44 &  66.17 &  55.97 &  1641.4 & {\bf 12.29}  & {\bf  204.21} \\
42.~ Singapore         & 69 &  82.24 &  62.29 &  1055.8 & {\bf  2.99}  & {\bf  197.32} \\
43.~ Brazil            & 43 &  62.14 &  48.59 &  1207.4 & {\bf 13.16}  & {\bf  171.11} \\
44.~ Iran              & 50 &  58.02 &  48.63 &   923.5 & {\bf  9.81}  & {\bf  152.83} \\
45.~ Romania           & 51 &  58.77 &  42.72 &   823.5 & {\bf  9.69}  & {\bf  142.67} \\
46.~ Lebanon           & 45 &  47.61 &  39.36 &  1077.9 & {\bf 12.23} & {\bf  140.86} \\
47.~ Brunei            & 70 &  61.24 &  38.97 &   740.4 & {\bf  1.70} & {\bf  137.23} \\ 
48.~ Ukraine           & 35 &  52.83 &  35.26 &   893.4 & {\bf 19.70} & {\bf  132.77} \\
49.~ Belarus           & 48 &  56.95 &  30.02 &   828.9 & {\bf 10.96} & {\bf  128.42} \\
50.~ South Africa      & 49 &  39.48 &  30.77 &   924.5 & {\bf 10.59} & {\bf  116.48} \\
\hline
\end{tabular}
\end{footnotesize} 
\caption{Like Table \ref{heplatrankingint}, but for the 7 {\tt arXiv}
sections. We see strong deviations between the ranks
for Georgia, Armenia, Estonia, Ukraine (e-rank $\ll \sigma$-rank),
and Singapore, Brunei, Norway, Austria, Malta, Australia
($\sigma$-rank $\ll$ e-rank).
Based on the e-rank, Moldova (37) and
Bosnia and Herzegovina (47) enter the top 50.}
\label{7arxivrankingint}
\end{table}

\label{sec:sum}
\section{Summary and Highlights}

We analyzed articles submitted to {\tt hep-lat} as primary
section in the period from 1992 to 2020. On average, there were
10.4 entries per week; after an initial increase, this has been
essentially stable since 1998. They were, on average, 25.8 times
cited (which is moderate compared other {\tt arXiv} section), but only
$46\,\%$ got published as regular papers. That is remarkably low
compared to other other {\tt arXiv} sections, most likely due to
the special role of the annual Lattice Conference proceedings.

There are 66 countries with lattice activities --- about a third
of the 195 countries worldwide --- in the sense that authors with
a working address there contributed to the {\tt hep-lat} section.

\begin{itemize}

\item If we consider the entire (extensive) national statistics,
these are the top 10 nations in the ranking based on the
Hirsch Index H: {\em USA, Germany, UK, Japan, France, Italy,
  Switzerland, Spain, Australia, Hungary.}\\
Since 2010, Germany caught up with the USA, and China is
significantly increasing its production of lattice articles
(in 2019 it attained $\Sigma$-rank 8, almost catching up with Spain).
We repeat that in the case of Switzerland we did not count articles
from CERN (if they do not involve another Swiss working address).

\item For comparison, if we sum over the 7 related {\tt arXiv} sections
listed in Section 2, the top 10 nations are:
{\em USA, France, Germany, UK, Italy, Spain, Russia, Canada, Japan,
Switzerland.}\\
This is similar to the {\tt hep-lat} top ten, but Russia and Canada
replace Australia and Hungary, {\it i.e.}\ the former are top in the
broader production, while the latter are more focused on the lattice.

\item If we consider the intensive production of {\tt hep-lat}
articles (relative to the population), the ranking changes
drastically. In that respect, the top 10 nations are:
{\em Cyprus, Switzerland, Ireland, Denmark, Germany, Austria, Hungary,
UK, Slovenia, Finland.}\\
Here six new nations appear, with relatively small populations.

\item In the intensive statistics over the 7 {\tt arXiv} sections,
again 5 new nations appear (compared to the extensive ranking)
in the top 10:
{\em Switzerland, Israel, Germany, Sweden, France, Denmark, Belgium,
UK, Finland, Italy.}
  
\end{itemize}  

It is noteworthy that the USA is leading in both extensive
statistics, but intensively it drops down to position 15 and
20, for {\tt hep-lat} and the 7 {\tt arXiv} sections,
respectively.

If we compare the intensive {\tt hep-lat} production to the GDPpp
or to the EI, we observe particularly high lattice activities in
{\em Cyprus, Hungary, Slovenia, Slovakia, Albania, Georgia, Armenia,
India} and {\em Bangladesh}. On the other hand, there are 129 nations
without any visible lattice activity.

Regarding the 7 {\tt arXiv} sections, {\em Armenia, Georgia, Kyrgyzstan}
and {\em Benin} have remarkably high physics activities
in light of their GDPpp. If we consider the EI instead, then
{\em Portugal, Benin, Morocco, India} and {\em Pakistan} have extraordinary
high activities in physics.

Of course, we can also express the
observations of Figure \ref{scat7arxiv} the other way round:
in light of the research activity in physics, the following
countries are remarkably wealthy: {\em Qatar, UAE, Kuwait, Luxembourg,
Brunei, Saudi Arabia, Oman, Singapore, Malaysia, Bahamas}
and {\em Norway.} Similarly, the level of general education is high
(with respect to physics activity) in {\em Sri Lanka, Latvia,
Philippines, Lithuania, New Zealand} and {\em Australia.}
Taking account of these different points of view,
we just present the facts, but we refrain from judging
these observations. \\

\noindent
{\bf Acknowledgement:} I thank the organizers of the XXXV International
Workshop on High Energy Physics ``From Quarks to Galaxies: Elucidating
Dark Sides'' --- in particular Vitaly Bornyakov --- for their kind
invitation.

\vspace*{-3mm}

\begin{thebibliography}{1}
\def\selectlanguageifdefined#1{
\expandafter\ifx\csname date#1\endcsname\relax
\else\selectlanguage{#1}\fi}
\providecommand*{\href}[2]{{\small #2}}
\providecommand*{\url}[1]{{\small #1}}
\providecommand*{\BibUrl}[1]{\url{#1}}
\providecommand{\BibAnnote}[1]{}
\providecommand*{\BibEmph}[1]{\emph{#1}}
\ProvideTextCommandDefault{\cyrdash}{\hbox to.8em{--\hss--}}
\providecommand*{\BibDash}{\ifdim\lastskip>0pt\unskip\nobreak\hskip.2em\fi
\cyrdash\hskip.2em\ignorespaces}

\bibitem{Wilson}
\selectlanguageifdefined{english}
\BibEmph{Wilson K.G.} {Confinement of Quarks}~//
\href{https://doi.org/10.1103/PhysRevD.10.2445}{Phys.\ Rev.\ D 10}
\BibDash
\newblock 1974. \BibDash
\newblock 2445--2459.

\bibitem{Polyakov}
\selectlanguageifdefined{english}
\BibEmph{Polyakov A.M.} {A View from the island}~//
{3rd International Symposium on the History of Particle Physics:
The Rise of the Standard Model}
\BibDash
\newblock 1992. \BibDash
\newblock 243--249. \BibDash
\newblock hep-th/9211140.

\bibitem{Urs}
\selectlanguageifdefined{english}
\BibEmph{Wenger U.} {From spin models to lattice QCD --
the scientific legacy of Peter Hasenfratz}~// {PoS LATTICE2016}
\BibDash
\newblock 2017. \BibDash
\newblock 024. \BibDash
\newblock arXiv:1702.00670 [hep-lat].
        
\bibitem{books}
\selectlanguageifdefined{english}
\BibEmph{Creutz, M.} {Quarks, Gluons and Lattices}.
\BibDash \newblock Cambridge University Press, 1983.
\BibEmph{Rothe, H.J.} {Lattice Gauge Theories: An Introduction}.
\BibDash \newblock World Scientific, 1992. 
\BibEmph{Montvay, I. and M\"{u}nster, G.} {Quantum Fields on a Lattice}.
\BibDash \newblock Cambridge University Press, 1994. 
\BibEmph{Smit, J.} {Introduction to Quantum Fields on a Lattice}.
\BibDash \newblock Cambridge University Press, 2002.
\BibEmph{DeGrand, T. and DeTar, C.}
{Lattice Methods for Quantum Chromodynamics}
\BibDash \newblock World Scientific, 2006.
\BibEmph{Gattringer, C. and Lang, C.B.}
{Quantum Chromodynamics on the Lattice}
\BibDash \newblock Springer, 2009.
\BibEmph{Knechtli, F., G\"{u}nther, M. and Peardon, M.}
{Lattice Quantum Chromodynamics: Practical Essentials}
\BibDash \newblock Springer, 2017.

\bibitem{QCD}
\selectlanguageifdefined{english}
\BibEmph{Bietenholz W. et~al.} {Flavour blindness and patterns of flavour symmetry breaking in lattice simulations of up, down and strange quarks}~//
\href{https://doi.org/10.1103/PhysRevD.84.054509}{Phys. Rev. D}
\BibDash
\newblock 2011. \BibDash
\newblock 05450. \BibDash
\newblock arXiv:1102.5300~[hep-lat].

\bibitem{Andreas}
\selectlanguageifdefined{english}
\BibEmph{Kronfeld A.S.} {Predictions with Lattice QCD}~//
\href{https://doi.org/10.1088/1742-6596/46/1/020}{J.\ Phys.\ Conf.\ Ser.\ 46}
\BibDash
\newblock 2006. \BibDash
\newblock 147--151. \BibDash
\newblock arXiv:hep-lat/0607011.
                
\bibitem{WBLat21}
\selectlanguageifdefined{english}
\BibEmph{Bietenholz W.} {The Evolution of Lattice Field Theory:
  a Statistical Study}~//
\href{https://doi.org/10.22323/1.396.0079}{Proc. Sci. (LATTICE2021)}
\BibDash
\newblock 2022. \BibDash
\newblock 079. \BibDash
\newblock arXiv:2110.01821~[hep-lat].

\bibitem{arXiv} https://arxiv.org/

\bibitem{Hirsch}
\selectlanguageifdefined{english}
\BibEmph{Hirsch J.E.} {An index to quantify an individual's
scientific research output}~//
\href{https://doi:10.1073/pnas.0507655102}{PNAS 102}
\BibDash
\newblock 2005. \BibDash
\newblock 16569--72. \BibDash
\newblock arXiv:physics/0508025.


\bibitem{UNO} United Nations Development Programme, http://hdr.undp.org/en/
(consulted in July 2020).


\end{thebibliography}

\end{document}